\documentclass[12pt]{amsart}

\usepackage{amssymb,amsthm,amsmath}
\usepackage[numbers,sort&compress]{natbib}
\usepackage{color}
\usepackage{graphicx}
\usepackage{amsthm}
\usepackage{tikz}
\usepackage{amssymb,amsthm,amsmath}
\usepackage[numbers,sort&compress]{natbib}
\usepackage{color}
\usepackage{graphicx}
\usepackage{tikz}
\usepackage{xparse}

\title[Bloch varieties   and quantum ergodicity ]{ Bloch varieties  and quantum ergodicity for periodic graph operators}

\author{Wencai Liu}
\address[W. Liu]{ Department of Mathematics, Texas A\&M University, College Station, TX 77843-3368, USA} \email{liuwencai1226@gmail.com; wencail@tamu.edu}



\theoremstyle{plain}
\newtheorem{theorem}{Theorem}[section]
\newcommand{\R}{\mathbb{R}}
\newtheorem{corollary}[theorem]{Corollary}
\newtheorem{lemma}[theorem]{Lemma}

\newtheorem{remark}{Remark}
\newcommand{\C}{\mathbb{C}}

\newcommand{\Z}{\mathbb{Z}}

\theoremstyle{plain}
\newtheorem{definition}{Definition}
\newtheorem{conjecture}{Conjecture}

\begin{document}
	
	
	\begin{abstract}
	For periodic 	graph operators, 
we establish criteria  to  determine the overlaps of spectral band functions based on   Bloch varieties.  One  criterion states that for a large family of  periodic 	graph operators, 
the irreducibility of Bloch varieties implies no non-trivial    periods for spectral band functions.
This particularly
shows that spectral band functions of  discrete periodic Schr\"odinger operators on $\mathbb{Z}^d$  have no non-trivial    periods, answering  a
  question asked by Mckenzie and  Sabri [Quantum ergodicity for periodic graphs. arXiv preprint arXiv:2208.12685 (2022)]. 

	\end{abstract}
	
	\maketitle 
	\section{Introduction and main results}
	
  Bloch and Fermi varieties  play a crucial role in the study of spectral theory of periodic Schr\"odinger operators and related models. We refer readers to a recent  review~\cite{ksurvey} which focuses  on techniques arising from Bloch and Fermi varieties.
 In the continuous setting,  Bloch and Fermi varieties are often analytic. 
 For discrete  periodic graph operators, both Bloch and Fermi varieties  are algebraic in appropriate coordinates.  
 
 
 Recently there are remarkable developments in using various tools such as algebraic methods,  techniques in  geometric combinatorics and theory in 
 complex analysis of multi-variables to study the (ir)reducibility of Bloch and Fermi varieties, Fermi isospectrality, density of states, and  critical points of spectral band functions of periodic graph operators ~\cite{flm,liu1,liujmp22,dks,shjst20,fls,lslmp20,sm22,ber,cos,liu2021fermi,liu11}.

 The main goal of this paper is 
   to use Bloch varieties to understand overlaps and periods of spectral band functions of periodic graph operators.  One of our motivations comes from a recent arxiv preprint  of Mckenzie and Sabri~\cite{ms}, where they  proved the quantum ergodicity for a family of periodic graph operators under  an  assumption on overlaps and  periods of the spectral band functions\footnote{In July 2022, Mostafa Sabri  asked me  a question (see Question 1 below) whether  or not  discrete periodic Schr\"odinger operators on $\Z^d$  satisfy the assumption, which he and Theo McKenzie need to establish quantum ergodicity. I got interested in this problem and finally wrote this paper. }.  As corollaries, we  give criteria to verify  for  which periodic graphs,   the assumption is satisfied. 
   
	
Our main results are general and independent of periodic graph operators.
Assume that  $\mathcal{A}=\mathcal{A}(z)$ is a $Q\times Q$ matrix  and each entry of $\mathcal{A}(z)$ is a Laurent polynomial of $z=(z_1,z_2,\cdots,z_d)\in(\C^\star)^d$, where $\C^\star=\C\backslash \{0\}$.  
	Let $z_j=e^{2\pi ik_j}$, $j=1,2,\cdots, d$ and $A(k)=\mathcal{A}(z)$. 
	
Obviously,  $A(k)$ is periodic with respect to $k$.  In the following, $\mathcal{A}(z)$ and $A(k)$ are always the same (with respect to different variables). 
	Assume that for  any $k\in\R^d$, $A(k)$  is Hermitian. Denote by $\lambda_A^j(k)$, $k\in\R^d, j=1,2,\cdots, d$,    eigenvalues of $ A(k)$ in the non-decreasing order:
	\begin{equation}\label{glab}
	\lambda_A^1(k)\leq \lambda_A^2(k)\leq \cdots\leq \lambda_A^Q(k),k\in \R^d.
	\end{equation}

For any $\eta=(\eta_1,\eta_2,\cdots,\eta_d)\in\C^d$ and $\zeta=(\zeta_1,\zeta_2,\cdots,\zeta_d)\in\C^d$,  let  $\eta \odot  \zeta=(\eta_1\zeta_1,\eta_2\zeta_2,\cdots,\eta_d\zeta_d)$.
Denote by $0_d$  and $1_d$ the zero vector and unit element in $\C^d$: $0_d=(0,0,\cdots,0)$ and $1_d=(1,1,\cdots,1)$.

\begin{definition}
	We say the spectral band functions of $A(k)$  have no non-trivial periods if the following statement holds.
	If 
	for some $ \alpha\in\R^d$  and    $s,w\in \{1,2,\cdots, Q\}$,  the set 
	\begin{equation}\label{Sg3}
	\{k\in\R^d:   \lambda_A^s(k+\alpha)=\lambda_A^w(k) \}
	\end{equation}
	has positive Lebesgue measure, then we must have 
  $\alpha=0_d\mod \Z^d$ and $s=w$.
\end{definition}

Since there are only finitely many choices of   $s,w\in \{1,2,\cdots, Q\}$,   the spectral band functions of $A(k)$  have no non-trivial periods if and only if 	for any $ \alpha \in\R^d$ with $\alpha\neq 0_d\mod \Z^d$, 
the set 
\begin{equation}\label{Sg4}
S_1(\alpha)=\{k\in\R^d: \text{ there exist } s \text{ and } w \text{ such that } \lambda_A^s(k+\alpha)=\lambda_A^w(k) \}
\end{equation}
has  Lebesgue measure zero, and 
the set 
\begin{equation}\label{Sg4new1}
S_2=\{k\in\R^d: \text{ there exist distinct } s \text{ and } w \text{ such that } \lambda_A^s(k)=\lambda_A^w(k) \}
\end{equation}
has Lebesgue measure zero.

%

Let
\begin{equation}
\mathcal{P}_{\mathcal{A}}(z,\lambda)=\det (\mathcal{A}(z)-\lambda I_{Q\times Q}).
\end{equation} 
Note that $\mathcal{P}_{\mathcal{A}}(z,\lambda)$ is  a Laurent polynomial in $z$ and a polynomial in $\lambda$. 
Let $\mathcal{P}_{\mathcal{A}}^l(z,\lambda)$, $l=1,2,\cdots,K$ be the non-trivial irreducible\footnote{Non-trivial  Laurent polynomials mean non-monomials.
} factors of $\mathcal{P}_{\mathcal{A}}(z,\lambda)$:
\begin{equation}\label{gire}
\mathcal{P}_{\mathcal{A}}(z,\lambda)=(-1)^Q\prod_{l=1}^K\mathcal{P}_{\mathcal{A}}^l(z,\lambda).
\end{equation}
It is easy to see that $\mathcal{P}_{\mathcal{A}}^l(z,\lambda)$ must depend on $\lambda$.
Since  $\mathcal{P}_{\mathcal{A}}(z,\lambda)$  is a polynomial in $\lambda$ with the highest degree term (in $\lambda$)   $(-1)^Q\lambda^Q$, we can normalize
 $\mathcal{P}_{\mathcal{A}}^l(z,\lambda)$ in the following way: 
  $\mathcal{P}_{\mathcal{A}}^l(z,\lambda)$ is a Laurent polynomial in $z$ and a polynomial in $\lambda$, and the coefficient of highest degree term of $\lambda$ is 1.

We say 
 $\mathcal{P}_{\mathcal{A}}(z,\lambda) $   is square-free if  for any distinct $l_1$ and $l_2$ in $\{1,2,\cdots,K\}$,
\begin{equation}\label{gsq}
\mathcal{P}^{l_1}_{\mathcal{A}}(z,\lambda)\not\equiv \mathcal{P}^{l_2}_{\mathcal{A}}(z,\lambda).
\end{equation}

Given $ \alpha\in\R^d$ with $\alpha\neq 0_d\mod \Z^d$,
we say   $\mathcal{P}_{\mathcal{A}}(z,\lambda) $ satisfies condition  $C_{\alpha}$ if for  
	any  $l_1$ and $l_2$ in $\{1,2,\cdots,K\}$,
	\begin{equation}\label{gdis7}
	\mathcal{P}^{l_1}_{\mathcal{A}}(z,\lambda)\not\equiv \mathcal{P}^{l_2}_{\mathcal{A}}(\zeta \odot z,\lambda),
	\end{equation}
where 
$\zeta=(e^{2\pi i \alpha_1},e^{2\pi i\alpha_2},\cdots, e^{2\pi i \alpha_d})$.

\begin{theorem}\label{thm1new}
	The following statements hold
	\begin{enumerate}
	
		\item $\mathcal{P}_{\mathcal{A}}(z,\lambda) $  satisfies $C_{\alpha}$ if and only if  $S_1(\alpha)$ has  Lebesgue measure zero;
			\item  $\mathcal{P}_{\mathcal{A}}(z,\lambda) $  is square-free  if and only if $S_2$ has  Lebesgue measure zero.
	\end{enumerate}
\end{theorem}
\begin{remark}
	From the proof of Theorem \ref{thm1new}, one can see that 
	\begin{itemize}
		\item for any $\alpha\in\R^d$, either   $S_1(\alpha)=\R^d$ or ${\rm Leb} (S_1(\alpha))=0$;
		\item either   $S_2=\R^d$ or ${\rm Leb} (S_2)=0$.
	\end{itemize}
	
\end{remark}
Theorem \ref{thm1} immediately implies 
\begin{corollary}\label{thm1}
Assume that  
$\mathcal{P}_{\mathcal{A}}(z,\lambda) $ is square-free,    and 
	for any $\zeta=(\zeta_1,\zeta_2,\cdots, \zeta_d)\in \C^d\backslash \{1_d\}$ with $ |\zeta_j|=1,j=1,2,\cdots,d$, and
	any  $l_1$ and $l_2$ in $\{1,2,\cdots,K\}$, 
	\begin{equation}\label{gdis7newnew}
	\mathcal{P}^{l_1}_{\mathcal{A}}(z,\lambda)\not\equiv \mathcal{P}^{l_2}_{\mathcal{A}}(\zeta \odot z,\lambda).
	\end{equation}
	Then the spectral band functions of $A(k)$ have no non-trivial periods.

\end{corollary}

Let $L_N^d = \{0,1,\cdots,N-1\}^d$. 
	\begin{corollary}\label{cor5}
	Given any $m = (m_1,m_2,\cdots,m_d) \in   L_N^d \backslash \{0_d\}$, let $\zeta(m,N)=(e^{2\pi i \frac{m_1}{N}},e^{2\pi i \frac{m_2}{N}},\cdots, e^{2\pi i \frac{m_d}{N}})$. 
Assume that   there exists $N_0$ such that for any $N\geq N_0$, any  $m   \in   L_N^d \backslash \{0_d\}$, and 
any  $l_1$, $l_2$ in $\{1,2,\cdots,K\}$, 
\begin{equation}\label{gdis7new}
\mathcal{P}^{l_1}_{\mathcal{A}}(z,\lambda)\not\equiv \mathcal{P}^{l_2}_{\mathcal{A}}(\zeta(m,N) \odot z,\lambda).
\end{equation}
Then    for any $s,w\in \{1,2,\cdots, Q\}$,  
\begin{equation}\label{maing1}
\lim_{N\rightarrow \infty}\sup_{\substack{m\in L_N^d\\m\neq 0_d}} \frac{\#\{r\in L_N^d : \lambda_A^s(\frac{r_j+m_j}{N})-\lambda_A^w(\frac{r_j}{N})=0\}}{N^d} = 0,
\end{equation}
where  $r = (r_1,r_2,\cdots,r_d)$. 
	\end{corollary}

Since irreducibility implies square-free, by Corollaries \ref{thm1} and \ref{cor5}, we have that 
	\begin{corollary}\label{cor3}
	Assume that $\mathcal{P}_{\mathcal{A}}(z,\lambda)$ is irreducible and for any $\zeta=(\zeta_1,\zeta_2,\cdots, \zeta_d)\in \C^d\backslash \{1_d\}$ with $ |\zeta_j|=1,j=1,2,\cdots,d$,
	\begin{equation}\label{gdis8}
	\mathcal{P}_{\mathcal{A}}(z,\lambda)\not\equiv \mathcal{P}_{\mathcal{A}}(\zeta \odot z,\lambda).
	\end{equation}
	Then the spectral band functions of $A(k)$ have no non-trivial periods and   for any $s,w\in \{1,2,\cdots, Q\}$,  
	\begin{equation*}
	\lim_{N\rightarrow \infty}\sup_{\substack{m\in L_N^d\\m\neq 0_d}} \frac{\#\{r\in L_N^d : \lambda_A^s(\frac{r_j+m_j}{N})-\lambda_A^w(\frac{r_j}{N})=0\}}{N^d} = 0.
	\end{equation*}
\end{corollary}
	\begin{theorem}\label{thm2}
	Assume $\mathcal{P}_{\mathcal{A}}(z,\lambda)$ is irreducible.
	Then for any $a\neq 0$ and   any  $ \alpha \in\R^d$, the set 
\begin{equation}\label{Sg4new}
\{k\in\R^d: \text{ there exist } s \text{ and } w \text{ such that } \lambda_A^s(k+\alpha)=\lambda_A^w(k)+a \}
\end{equation}
has  Lebesgue measure zero. 
\end{theorem}
In \cite{ms},  under the  assumption   \eqref{maing1}, Mckenzie and Sabri proved the quantum ergodicity for periodic  graph operators.    Roughly speaking,  quantum ergodicity means that most eigenfunctions  on periodic graphs are equidistributed. We refer readers to ~\cite{ms} for the precise description of quantum ergodicity. 

Now we want to discuss the applications of   our main results to quantum ergodicity. In this paper, we  focus on 
Corollary  \ref{cor3}. The  requirement  \eqref{gdis8} is   easy to verify.  So the only   restriction of   applying Corollary  \ref{cor3} is the irreducibility of $\mathcal{P}_{\mathcal{A}}(z,\lambda)$.  
 In applications, starting with a periodic graph operator $H$ and Floquet-Bloch boundary condition (depends on $k\in\R^d$ or $z\in(\C^{\star})^d$), we obtain a matrix $A(k)$ ($\mathcal{A}(z)$). 
The irreducibility of $\mathcal{P}(z,\lambda)$ in  Corollary  \ref{cor3}  essentially means the irreducibility of the Bloch variety of $H$ (modulo periodicity): 
\begin{equation}
B=\{(k,\lambda)\in\C^{d+1}: z_j=e^{2\pi i k_j}, \mathcal{P}_{\mathcal{A}} (z,\lambda)=0\}.
\end{equation}

Recently, the author applied algebraic methods to obtain more general
proofs of irreducibility for Laurent polynomials, including  proving  the irreducibility of Bloch and Fermi varieties (Fermi variety is the level set of the Bloch variety) for discrete periodic Schr\"odinger operators in arbitrary dimension~\cite{liu1}, which previously were only studied in two and three dimensions ~\cite{battig1988toroidal,GKTBook,batcmh92,ktcmh90,bktcm91}. The approach in~\cite{liu1}  has been developed by Fillman, Matos and the author to prove the irreducibility of Bloch varieties for  a large family of  periodic graph operators~\cite{flm}.
So Corollary  \ref{cor3} may be applicable to many models.
In the following,  we only discuss one  case:   discrete periodic Schr\"odinger operators on $\Z^d$.  
In ~\cite{ms} Mckenzie and Sabri  asked a question: 

{\bf Question 1:} Do discrete periodic Schr\"odinger operators on $\Z^d$  satisfy the assumption   \eqref{maing1} ?

As an application of Corollary  \ref{cor3}, we answer Question 1 positively.



Let us give the precise definition of discrete periodic Schr\"odinger operators on $\Z^d$.
Given positive integers  $q_j$, $j=1,2,\cdots,d$,
let $\Gamma=q_1\Z\oplus q_2 \Z\oplus\cdots\oplus q_d\Z$.
We say that a function $V: \Z^d\to \R$ is  $\Gamma$-periodic (or just periodic)  if 
for any $\gamma\in \Gamma$,  $V(n+\gamma)=V(n)$.

Let  $\Delta$ be the discrete Laplacian on $\Z^d$, namely
\begin{equation*}
(\Delta u)(n)=\sum_{||n^\prime-n||_1=1}u(n^\prime),
\end{equation*}
where $n=(n_1,n_2,\cdots,n_d)\in\Z^d$, $n^\prime=(n_1^\prime,n_2^\prime,\cdots,n_d^\prime)\in\Z^d$ and 
\begin{equation*}
||n^\prime-n||_1=\sum_{i=1}^d |n_i-n^\prime_i|.
\end{equation*}
Consider the  discrete    Schr\"{o}dinger operator on $ \Z^d$,
\begin{equation} \label{h0}
H=\Delta +V ,
\end{equation}
where $V$ is  $\Gamma$-periodic.

	Let $\{\textbf{e}_j\}$, $j=1,2,\cdots d$,  be the   standard basis in $\Z^d$:
	\begin{equation*}
	\textbf{e}_1 =(1,0,\cdots,0),\textbf{e}_2 =(0,1,0,\cdots,0),\cdots, \textbf{e}_{d}=(0,0,\cdots,0,1).
	\end{equation*}

		Let us consider  the equation
		\begin{equation} 
		(\Delta u)(n)+V(n)u(n)=\lambda u(n) \label{spect_0}, n\in\Z^d,
		\end{equation}
		with the so called Floquet-Bloch boundary condition
		\begin{equation}  
		u(n+q_j\textbf{e}_j)=z_j u(n)=e^{2\pi i k_j}u(n),j=1,2,\cdots,d, \text{ and } n\in \Z^d.\label{Fl}
		\end{equation}


		Let $D_V (k)$   ($\mathcal{D}_V(z)$) be the periodic operator $\Delta+V$ with the Floquet-Bloch boundary condition \eqref{Fl} with respect to variables $k$ ($z$). $D_V (k)$  can be realized as  a $Q\times Q$ matrix, where $Q=q_1q_2\cdots q_d.$
	 Let $\lambda_V^j(k)$, $j=1,2,\cdots, Q$ be the standard spectral band functions of $\Delta +V$ (applying \eqref{glab} with $A(k)=D_V(k)$).  
		
		\begin{corollary}\label{cor1}
			For any discrete periodic Schr\"odinger operators $\Delta+V$, 
			we have that 
		  for any $s,w\in \{1,2,\cdots, Q\}$,  
			\begin{equation}\label{maing2}
			\lim_{N\rightarrow \infty}\sup_{\substack{m\in L_N^d\\m\neq 0_d}} \frac{\#\{r\in L_N^d : \lambda_V^s(\frac{r_j+m_j}{N})-\lambda_V^w(\frac{r_j}{N})=0\}}{N^d} = 0.
			\end{equation}
		\end{corollary}
	
	\begin{remark}
	 Corollary \ref{cor1} answers Question 1 positively.

	\end{remark}

%
\section{Proof of  Theorem \ref{thm1new}, Corollary \ref{cor5} and  Theorem \ref{thm2}}
%
%
%

\begin{proof}[\bf Proof of Theorem \ref{thm1new}]
Recall that  $\alpha\neq 0_d\mod \Z^d$, $ \zeta_j=e^{2\pi i \alpha_j} $ and  $z_j=e^{2\pi i k_j},j=1,2,\cdots,d$.
Let 
\begin{equation*}
P_A(k,\lambda)= \mathcal{P}_{\mathcal{A}} (z,\lambda)=\det (A(k)-\lambda I).
\end{equation*}
Note that $P_A(k,\lambda)$ is analytic.

If $\mathcal{P}_{\mathcal{A}}$ does not satisfy $C_{\alpha}$,  then 
there exist   $l_1$ and $l_2$ in $\{1,2,\cdots,K\}$ ($l_1$ may equal $l_2$) such that 
\begin{equation}\label{gdis71}
\mathcal{P}^{l_1}_{\mathcal{A}}(z,\lambda)\equiv \mathcal{P}^{l_2}_{\mathcal{A}}(\zeta \odot z,\lambda).
\end{equation}
This implies that for any $k$, $A(k)$ and $A(k+\alpha)$ have at least one common eigenvalue and hence there exist 
$ s, w$ such that $\lambda_A^s(k+\alpha)=\lambda_A^w(k)$.  Therefore, $S_1(\alpha)=\R^d$.

Simple calculations imply that 
\begin{align}
{S}_1(\alpha)&=\{k\in\R^d: \text{ there exist } s, w \text{ such that } \lambda_A^s(k+\alpha)=\lambda_A^w(k) \}\nonumber \\
&=\{k \in\R^d: \text{ and there exists } \lambda \text{ such that }\nonumber\\
&\;\;\;\;\;\; P_A(k,\lambda)= P_{A}(\alpha+  k,\lambda)=0 \}\nonumber\\
&= \text{ Proj}_k\{(k,\lambda) \in \R^{d+1}:  P_A(k,\lambda)= P_{A}(\alpha+  k,\lambda)=0  \},\label{g2new}
\end{align}
where $\text{ Proj}_k$ is the projection to $k$ variables.

If  $\mathcal{P}_{\mathcal{A}}$   satisfies $C_{\alpha}$,  one has that 
the algebraic variety 
$ \{(z,\lambda):\mathcal{P}_{\mathcal{A}}(z,\lambda)= \mathcal{P}_{\mathcal{A}}(\zeta\odot z,\lambda)=0\}$ has algebraic dimension $d-1$.   This implies that $\{(k,\lambda) \in \R^{d+1}:  P_A(k,\lambda)= P_{A}(\alpha+  k,\lambda)=0  \}$ has real (analytic) dimension at most $d-1$. By   \eqref{g2new},  one has that 
 ${S}_1(\alpha)$   (as a subset in $\R^d$) has Lebesgue measure zero.	We finish the proof of  part 1.

If $\mathcal{P}_{\mathcal{A}}$ is not square-free,  namely there exist 
 distinct $l_1$ and $l_2$ in $\{1,2,\cdots,K\}$ such that 
\begin{equation}\label{gsqnew}
\mathcal{P}^{l_1}_{\mathcal{A}}(z,\lambda)\equiv \mathcal{P}^{l_2}_{\mathcal{A}}( z,\lambda).
\end{equation}
This implies that 
  for any $k$, $A(k)$  has repeated eigenvalues and hence there exist distinct 
$ s, w\in\{1,2,\cdots, Q\}$  such that $\lambda_A^s(k)=\lambda_A^w(k)$.  Therefore, $S_2=\R^d$.

It is clear that 
\begin{align}
 {S}_2&=\{k\in\R^d: \text{ there exist distinct } s, w \text{ such that } \lambda_A^s(k)=\lambda_A^w(k) \} \nonumber\\
&=\{k\in \R^d:  A(k) \text{ has repeated eigenvalues } \}\nonumber\\
&=\{k\in \R^d:  \text{there exists } \lambda \text{ such that } P_A(k,\lambda)= \partial_{\lambda} P_A(k,\lambda)=0 \}\nonumber\\
&= \text{ Proj}_k\{(k,\lambda)\in\R^{d+1}:  P_A(k,\lambda)= \partial_{\lambda} P_A(k,\lambda)=0 \}.\label{g10}
\end{align}

If $\mathcal{P}_{\mathcal{A}}$ is square-free, then 
the algebraic variety 
$ \{(z,\lambda):\mathcal{P}_{\mathcal{A}}(z,\lambda)= \partial_{\lambda}\mathcal{P}_{\mathcal{A}}( z,\lambda)=0\}$ has algebraic dimension $d-1$. This implies that $\{(k,\lambda)\in\R^{d+1}:  P_A(k,\lambda)= \partial_{\lambda} P_A(k,\lambda)=0 \}$ 
has real (analytic) dimension at most $d-1$. By   \eqref{g10},  one has that 
${S}_2$   (as a subset in $\R^d$) has Lebesgue measure zero.
We finish the proof of part 2.

	\end{proof}
\begin{proof}[\bf Proof of Corollary \ref{cor5}]
Let 
\begin{equation*}
U=\bigcup_{N\geq N_0}\bigcup_{\substack{m \in L_N^d\\ m\neq 0_d}} {S}_1\left({\frac{m}{N}}\right). 
\end{equation*}
Applying  Part 1 of Theorem \ref{thm1new} with all $\alpha=\frac{m}{N}$ with $m \in L_N^d\backslash \{0_d\}$ and $N\geq N_0$,  
$U$ has Lebesgue measure zero.
Therefore,  for any $s,w\in \{1,2,\cdots, Q\}$ one has that 
\begin{align}
\lim_{N\rightarrow \infty}\sup_{\substack{m\in L_N^d\\m\neq 0_d}} \frac{\#\{r\in L_N^d : \lambda_A^s(\frac{r_j+m_j}{N})-\lambda_A^w(\frac{r_j}{N})=0\}}{N^d} &\leq \lim_{N\rightarrow \infty}\frac{1}{N^d} \sum_{r\in L_N^d}  \chi_{U} \left(\frac{r_j}{N}\right)\\
&=0,\label{g3}
\end{align}
where $\chi_{U} $ is the characteristic function.  
\end{proof}
\begin{proof}[\bf Proof of Theorem \ref{thm2} ]
 From the proof of  part 1 of Theorem \ref{thm1new}, it suffices to show that 
  for any $a\neq 0$ and any $\zeta \in \C^d $ with $ |\zeta_j|=1,j=1,2,\cdots,d$, 
 \begin{equation}\label{gdis6}
 \mathcal{P}_{\mathcal{A}}(z,\lambda)\not\equiv \mathcal{P}_{\mathcal{A}}(\zeta \odot z,\lambda+a).
 \end{equation}
 Simple calculations imply that
 \begin{equation}\label{g7}
 \mathcal{P}_{\mathcal{A}}(z,\lambda)=(-\lambda)^Q+\text{Tr} \mathcal{A}(z) (-\lambda)^{Q-1}+\text{l.o.t}
 \end{equation}
 and 
  \begin{equation}\label{g8}
  \mathcal{P}_{\mathcal{A}}(\zeta \odot  z,\lambda+a)=(-\lambda)^Q+(-Qa+\text{Tr} \mathcal{A}(\zeta \odot  z) )(-\lambda)^{Q-1}+\text{l.o.t},
 \end{equation}
 where $ \text{l.o.t}$ contains terms of $\lambda$ with degree less than or equal to $Q-2$. 
 Obviously, constant terms in both $\text{Tr} \mathcal{A}(z)$ and $\text{Tr} \mathcal{A}(\zeta \odot  z)$ are the same.  Then 
 $\text{Tr} \mathcal{A}(z) $ and $-Qa+\text{Tr} \mathcal{A}(\zeta \odot  z)$ are different functions.  
 Now \eqref{gdis6} follows from \eqref{g7} and \eqref{g8}.
\end{proof}

\section{Proof of Corollary \ref{cor1}}
In this section, we first recall some basics. We refer readers to \cite{liu1} for details.
For $n=(n_1,n_2,\cdots, n_d)$, let $z^n = z_1^{n_1}\cdots z_d^{n_d}$.
By abusing the notation, denote by $q=(q_1,q_2,\cdots, q_d)$.
Let  $\hat{V}(n)$,  $n \in \Z^d$   be the discrete Fourier transform of $\{V(n)\}$.

Define
\begin{equation}\label{gtm}
\tilde{\mathcal{D}}_V(z)= \mathcal{D}_V(z^q),
\end{equation}
and 
\begin{equation}\label{gtp}
\tilde{\mathcal{P}}_V(z,\lambda)=\det( \tilde{\mathcal{D}}_V(z,\lambda)-\lambda I)= \mathcal{P}_V(z^q,\lambda).
\end{equation}
Let $$\rho^j_{n_j}=e^{2\pi i \frac{n_j}{q_j}},$$
where $0\leq n_j \leq q_j-1$, $j=1,2,\cdots,d$.

By the standard discrete Floquet transform (e.g.,  \cite{liu1,ksurvey}), one has
\begin{lemma}\label{lesep} 
	$\tilde{\mathcal{D}}_V(z)$ is unitarily equivalent to 		
	$
	B_0+B_V,
	$
	where $B_0$ is a diagonal matrix with entries
	\begin{equation}\label{A}
	B_0(n;n^\prime)=\left(\sum_{j=1}^d \left(\rho^j_{n_j}z_j+\frac{1}{\rho^j_{n_j} z_j} \right)\right) \delta_{n,n^{\prime}}
	\end{equation}
	 \begin{equation}\label{gb}
	B_V(n;n^\prime)=\hat{V} \left(n_1-n_1^\prime,n_2-n_2^\prime,\cdots, n_d-n_d^\prime\right),
	\end{equation}
	and
	\begin{equation*}
0\leq n_j\leq q_j-1, 0\leq n_j'\leq q_j-1, j=1,2,\cdots,d.
	\end{equation*}
	In particular,
	\begin{equation*}
	\tilde{\mathcal{P}}_V(z, \lambda) =\det(B_0+B_V-\lambda I).
	\end{equation*}

\end{lemma}
%

Let 
	\begin{equation}\label{H}
h(z,\lambda) = \prod_
{\substack {0\leq n_j \leq q_j-1\\ 1\leq j\leq d} }((\sum_{j=1}^d\rho_{n_j}^j z_j)-\lambda). 
\end{equation}
\begin{proof}[\bf Proof of Corollary \ref{cor1}]
	Recall that $\mathcal{P}_{V}(z,\lambda)$ is irreducible \cite{liu1,flm}.
By Corollary \ref{cor3}, it suffices to verify that for any $\zeta=(\zeta_1,\zeta_2,\cdots, \zeta_d)\in \C^d\backslash \{1_d\}$ with $ |\zeta_j|=1,j=1,2,\cdots,d$,
\begin{equation}\label{gdis1}
\mathcal{P}_{V}(z,\lambda)\not \equiv \mathcal{P}_{V}(\zeta \odot z,\lambda).
\end{equation}
It suffices to prove that  for any $\zeta=(\zeta_1,\zeta_2,\cdots, \zeta_d)\in \C^d  $ with $ |\zeta_j|=1,j=1,2,\cdots,d$ and $\zeta^q\neq1_d$, one has that
\begin{equation}\label{gdis2}
\tilde{\mathcal{P}}_{V}(z,\lambda)\not \equiv \tilde{\mathcal{P}}_{V}(\zeta \odot z,\lambda).
\end{equation}
By \eqref{H} and Lemma \ref{lesep},  $h(z,\lambda)$ is the  highest  degree component of $\tilde{\mathcal{P}}_{V}(z,\lambda)$.
Therefore, to prove  \eqref{gdis2}, 
it suffices to show that  for any $\zeta=(\zeta_1,\zeta_2,\cdots, \zeta_d)\in \C^d  $ with $ |\zeta_j|=1,j=1,2,\cdots,d$ and $\zeta^q\neq 1_d$, one has that
\begin{equation}\label{gdis3}
h(z,\lambda)\not \equiv h(\zeta \odot z,\lambda).
\end{equation}
It is easy to see that \eqref{gdis3} holds.

\end{proof}
	\section*{Acknowledgments}
	
	This research was supported by NSF DMS-2000345 and  DMS-2052572.  
  I would like to express my sincere gratitude to Mostafa  Sabri for drawing me attention to   Question 1  (see footnote 1),  many valuable  discussions  on this subject and comments on earlier versions of this paper. 

	
\end{document}